%
%
%
%
%
%
%
%
\input amstex
\input amsppt.sty



\NoBlackBoxes
\topmatter
\title     
Coherent states, Entanglement, and Geometric Invariant Theory
\endtitle
\leftheadtext{Entanglement} 
\rightheadtext{A.~Klyachko}
\author
Alexander A.~Klyachko
\endauthor
\address
Bilkent University \newline\indent Bilkent, 06533 Ankara Turkey
\endaddress
\email klyachko\,\@\,fen.bilkent.edu.tr
\endemail
%
%
\affil Bilkent University, Ankara, Turkey
\endaffil
\toc
\head
Introduction
\endhead
\head
1. Coherent states
\endhead
\subhead
1.1. Glauber's coherent states
\endsubhead
\subhead
1.2. Dynamic group and general coherent states
\endsubhead
\subhead
1.3. Total variance and extremal property
\endsubhead
\head 2. Entanglement
\endhead
\subhead
2.1. EPR paradox
\endsubhead
\subhead
2.2. Bell's paradox
\endsubhead
\subhead
2.3. Ansatz for testing ``classical realism"
\endsubhead
\subhead
2.4. Extremal property of completely entangled states
\endsubhead
\subhead
2.5. Formal definition and examples
\endsubhead
\subhead
2.6. Kempf-Ness unitary trick and GIT stability
\endsubhead
\subhead
2.7. Density matrix and measure of entanglement
\endsubhead
\subhead
2.8. Hilbert-Mumford criterion
\endsubhead
\head
3. Conclusion
\endhead
\head
References
\endhead
\endtoc
\endtopmatter
\document

\head  Introduction
\endhead
From a thought experiment for testing the very basic principles of
quantum mechanics in its early years \cite{EPR35, Schr\"o35},
entanglement nowadays is growing into an important technical tool
for quantum information processing \cite{EPR96,...}. Surprisingly
enough currently there is no agreement of opinion among  experts
on the very definition of entanglement, and its proper measure
\cite{Peres98, BarPho89, PleVed98, VedPle98, BZZ01, BZ01}. Here we
propose a new approach to entanglement, based on dynamic symmetry
group of a quantum system. A similar approach was applied by
A.~Perelomov \cite{Perel86} to coherent states, which in many
respects are opposite to entangled ones. The celebrated
``unexpected efficiency" \cite{Wig67} of group-theoretical methods
in quantum mechanics was many times demonstrated by E.~Wigner
\cite{Wig31, Wig39}, whose centenary holds this year.

The main objective of the paper is to unveil an adequate
mathematics hidden behind entanglement, that is {\it Geometric
Invariant Theory} \cite{MFK94}. More specifically relation between
these two subjects can be described by the following theses.
\roster
\item Total variance of {\it completely entangled} state $\psi$ is maximal.
\item This distinguishes $\psi$ as a minimal vector
in its orbit under action of {\it complexified} dynamic group
$G^c$.
\item An ultimate aim of Geometric Invariant Theory is
a description of complex orbits and their minimal vectors.
It suggests that {\it noncompletely} entangled states are just GIT
semistable vectors.
\endroster
This approach provides a powerful tool for treatment  of
entanglement, and shed new light on some old problems. We consider
many classical and not so classical examples in support of these
theses, and discuss their relation with conventional approach.
Formal proofs are mostly skipped, and will be published elsewhere,
since they help little in search for a proper definition of an
unclear physical concept.

I first start think about the subject in Erwin Schr\"odinger
Institute of Mathematical Physics in Vienna in January 2001, and
would like  to express here my gratitude for financial support and
an exiting atmosphere.

This paper arouses from an attempt to unveil a hidden meaning of
some words
used by physicists. The following quotation from A.~Grothendieck
exposes the encountered difficulties
\roster\item""{\it
``Passer de la m\'echanique de Newton \`a celle d'Einstein doit
\^etre un peu, pour le math\'ematicien, comme de passer du bon
vieux dialecte proven\c{c}al \`a l'argot parisien dernier cri. Par
contre, passer a la m\'echanique quantique, j'imagine, c'est
passer du fran\c{c}ais au chinois."}
\footnote{To pass from Newton's mechanics  to that of Einstein
must be as easy, for mathematician, as to pass from good old
provincial dialect to the last cry of Paris slang. On the
contrary, to pass to the quantum mechanics, I think, is to pass to
Chinese.} \quad
\endroster

\head 1.~Coherent states\endhead
Coherent states, first introduced by Schr\"odinger
\cite{Schr\"o26}, lapsed into obscurity for decades until Glauber
\cite{Glaub63} rediscovered them in connection with laser
emission. Later on Perelomov \cite{Perel86} put them into an
adequate context of dynamic symmetry group. We'll use a similar
approach for entanglement, and to warm up recall here some basic
facts about coherent states.

\subhead
1.1.~Glauber's coherent states
\endsubhead
Let's start with {\it quantum oscillator}, described by canonical
pair of operators $p$, $q$,  $[p,q]=i\hbar$, generating  {\it
Weil-Heisenberg algebra} $\Cal W$. This algebra has unique unitary
irreducible representation, which can be realized in {\it Fock
space} $\Bbb F$ spanned by orthonormal set of $n$-excitations
states $|n\rangle$ on
 which dimensionless annihilation and creation operators
$$a=\frac{q+ip}{\sqrt{2\hbar}},\quad
a^\dag=\frac{q-ip}{\sqrt{2\hbar}}, \quad [a,a^\dag]=1$$ act by
formulae
$$a|n\rangle=\sqrt{n}|n-1\rangle, \quad
a^\dag|n\rangle=\sqrt{n+1}|n+1\rangle.$$
A typical element from
{\it Weil-Heisenberg group} $W=\exp\Cal W$, up to a phase factor,
is of the form $D(\alpha)=\exp(\alpha a^\dag -\alpha^* a)$ for
some $\alpha\in\Bbb C$. Action of this operator on vacuum
$|0\rangle$ produces state
$$|\alpha\rangle:=D(\alpha)|0\rangle=
\exp\left(-\frac{|\alpha|^2}{2}\right)
\sum_{n\ge 0}\frac{\alpha^n}{\sqrt{n!}}|n\rangle,$$
known as {\it Glauber coherent state}. The number of excitations
in this state has Poisson distribution with parameter
$|\alpha|^2$. In many respects its behavior  is close to classical
\cite{Perel86}, e.g. Heisenberg's uncertainty $\Delta p\Delta
q=\hbar/2$ for this state is minimal. We can summarize this
construction as follows:
$$\text{\it Glauber's coherent states} = \text{\it $W$-orbit of vacuum}.\tag1.1$$
\subhead 1.2.~Dynamic group and general coherent states
\endsubhead
Let's now turn to arbitrary quantum system $S$ with dynamic
symmetry group $G=\exp\Cal G$. By definition its  Lie algebra
$\Cal G$ is generated by all {\it essential observables\,} of the
system
\footnote{To eliminate irrelevant phase factors we expect that the
dynamic symmetry group acts on the state space by {\it
unimodular\,} transformations. Equivalently, Lie algebra of
observables  consists of {\it traceless} operators.} (like $p,q$
in the above example). To simplify the underling mathematics
suppose in addition that state space $\Bbb H=\Bbb H(S)$ of the
system is finite, and representation of $G$ in $\Bbb H$ is
irreducible.

To extend (1.1) to this general setting  we have to understand the
special role of the vacuum, which primary considered as a {\it
ground state} of a system.
For group-theoretical approach, however, another its property is
more relevant:
$$\text{\it Vacuum is a state with maximal symmetry.}\tag1.2$$
This may be also spelled out that vacuum is a most degenerate
state of a system. Symmetries of state $\psi$ are given by its
{\it stabilizers}
$$G_\psi=\{g\in G\mid g\psi=\lambda\psi\},\quad {\Cal
G}_\psi=\{X\in\Cal G\mid X\psi=\lambda\psi\}\tag1.3$$ in the
dynamic group $G$ or in its Lie algebra $\Cal G$. Looking back to
the quantum oscillator, we see that some symmetries are actually
hidden, and manifest themselves only in {\it complexified} algebra
${\Cal G}^c=\Cal G\otimes\Bbb C$ and group $G^c=\exp\Cal G^c$. For
example, stabilizer of vacuum in Weyl algebra $\Cal W$ consists of
scalars,
while in complexified algebra $\Cal W^c$ it contains a nonscalar
annihilation operator, $\Cal W_0^c=\Bbb C+\Bbb Ca$. In the last
case the stabilizer is big enough to recover the whole dynamic
algebra
$$\Cal W^c=\Cal W^c_{0}+{\Cal W^c_0}^\dag.$$
This decomposition, called {\it complex polarization}, gives a
precise meaning for the maximal degeneracy of a vacuum or a
coherent state \cite{Perel86}. It ensures that dimension of the
symmetry group of such state is at least half of dimension the
whole dynamic group.
\proclaim{1.2.1.~Definition} State $\psi\in \Bbb H$ is said to be coherent
\footnote{This is what Perelomov called ``coherent state closest
to classical". His generalized coherent states are defined as
elements  from $G$-orbit of an {\it arbitrary\,} initial vector
$\psi_0$. They have no intrinsic meaning, and are useful mainly as
a calculation tool.}  if
$$\Cal G^c=\Cal G^c_\psi+{\Cal G^c_\psi}^\dag,\tag1.4$$
where  ${\Cal G^c_\psi}^\dag$ consists of operators conjugate to
that of stabilizer ${\Cal G^c_\psi}$.
\endproclaim
In finite dimensional case all such decompositions come from {\it
Borel subalgebra}, i.e. a maximal solvable subalgebra $\Cal
B\subset\Cal G^c$. Algebra of all upper triangular matrices $\Cal
T\subset
\Cal G=
\text{Mat}(n,\Bbb C)$ is a typical example.
It is a basic structural  fact that
$\Cal B+\Cal B^\dag=\Cal G^c$, and  therefore
$$\psi \text{ is a coherent state}\quad\Longleftrightarrow \quad\psi
\text{ is an eigenvector of a Borel subalgebra }\Cal B.\tag1.5$$
In representation theory eigenstate $\psi$ of $\Cal B$  is called
{\it highest vector}, and
the corresponding eigenvalue $\lambda=\lambda(X)$,
$$X\psi=\lambda(X)\psi,\quad X\in \Cal B\tag1.6$$
is said to be {\it highest weight}. Here are their basic
properties:
\roster\item
For irreducible space $\Bbb H$ the highest vector $\psi_0$
(=vacuum) is unique.
\item
There is only one irreducible representation $\Bbb H=\Bbb
H_\lambda$ with highest weight $\lambda$.
\item
All coherent states are of the form $\psi=g\psi_0$, $g\in G$ (cf.
with (1.1)).
\item
Coherent state $\psi$ in composite system $\Bbb H=\Bbb
H_1\otimes\Bbb H_2$ splits into product $\psi=\psi_1\otimes
\psi_2$ of coherent states of the components.
\endroster
\subsubhead{1.2.2.~Remark}\endsubsubhead
One can spell out these properties by saying that unitary
irreducible representations of group $G$ are parameterized by
symmetry type of their coherent states or vacua. Coherent state
theory, in the form given by Perelomov \cite{Perel86}, is a
physical equivalent of Kirillov--Kostant {\it orbit method} in
representation theory \cite{Kiril76}.

Notice that in many cases the complexified symmetry group is
physically meaningful.
\subsubhead1.2.3.~Example\endsubsubhead
For a particle of spin $j>0$ the dynamic symmetry group is
$\text{SU}(2)$. Its complexification is group of unimodular
matrices $\text{SL}(2,\Bbb C)$, which, as first noted by Wigner
\cite{Wig39}, is a double cover of {\it Lorentz group}. It
is responsible for relativistic transformation of spin state in a
moving frame \cite{PST02}.

Coherent states in this example are those with definite spin
projection $j$ onto some direction. Group of {\it complex}
symmetries of such state is conjugate to group of triangular
matrices (=Borel subgroup).

\subhead 1.3.~Total variance and extremal property\endsubhead
Let us define {\it total variance} of state $\psi$ by equation
$${\Bbb D}(\psi)=\sum_i
\langle\psi|X_i^2|\psi\rangle-\langle\psi|X_i|\psi\rangle^2,\tag1.7$$
where $X_i$ form an orthonormal basis in Lie algebra $\Cal G$ with
respect to its invariant metric (for spin group $\text{SU}(2)$ one
can take moment operators  $J_x$, $J_y$, $J_z$ as $X_i$). The
total variance is $G$-invariant $\Bbb D(g\psi)=\Bbb D(\psi)$, $
g\in G,$ and independent of the basis $X_i$. It measures the total
level of {\it quantum fluctuations\,} of a system in state $\psi$.

The first sum in (1.7) contains well known {\it Casimir operator}
$$C=\sum_i X_i^2,\tag1.8$$
which acts as a scalar in every irreducible representation $\Bbb
H$ of $G$. For spin $j$ representation $\Bbb H_j$ of
$\text{SU}(2)$ the Casimir is equal to square of moment $j(j+1)$,
and in general $C=\langle\lambda,\lambda+\rho\rangle$ in
representation $\Bbb H_\lambda$ with highest weight $\lambda$
(here we use H.~Weyl's notation $\rho$ for halfsum of positive
roots). Hence
$$\Bbb D(\psi)=\langle\lambda,\lambda+\rho\rangle-
\sum_i\langle\psi|X_i|\psi\rangle^2.\tag1.9$$

\proclaim{1.3.1.~Theorem} State $\psi$ is coherent iff its
total variance is minimal, and in this case
$$\Bbb D(\psi)=\langle\lambda,\rho\rangle.\tag1.10$$
\endproclaim
This theorem, in a slightly less precise form, belongs to Delbargo
and Fox \cite{DelFox70}. It supports a common believe, that
coherent states are closest to classical ones. Note however that
such simple characterization holds only for finite dimensional
systems. The total variance, for example, makes no sense for
quantum oscillator, for which we have {\it minimality of
uncertainty} $\Delta p\Delta q=\hbar/2$ instead.
\subsubhead 1.3.2.~Example\endsubsubhead Recall that for spin $j$
representation of $\text{SU}(2)$ coherent state $\psi$ has
definite spin projection $j$ onto some direction, and by (1.10)
$\Bbb D(\psi)=j$. The standard deviation $\sqrt{j}$ for such state
is of smaller order then $j$,  therefore for $j\rightarrow
\infty$ it behaves classically  \cite{Perel86}.

\head
2. Entanglement
\endhead
Everybody knows, and nobody understand what is entanglement. The
very term  was coined in the famous Schr\"odinger's ``cat paradox"
\footnote{As BBC puts it: {\it In quantum mechanic it is not so
easy to be or not to be.}}
 paper \cite{Schr\"o35}, which in
turn was  inspired by the no less celebrated
Einstein--Podolsky--Rosen gedanken  experiment \cite{EPR35}. While
the authors were amazed by nonlocal nature of correlations between
involved particles, J.~Bell was the first to note that the
correlations themselves, put aside the nonlocality, are
inconsistent with ``classical realism" \cite{Bell65}. Since then
Bell's inequalities are produced in industrial quantities
\cite{CHSH69, GHZ89, GHSZ90, Merm90, WerWol01,...}. Neither of
this effects, however, allows decisively distinguish entangled
states from others.
Therefore we develop another approach, based on the dynamic
symmetry group.

\subhead 2.1.~EPR paradox\endsubhead
Decay of a spin zero state into two components of spin 1/2
subjects to a strong correlation between spin projections of the
components,
caused by conservation of moment. The correlation apparently
creates an {\it information channel\,} between the components,
acting beyond their light cones. This paradox, recognized in early
years of quantum mechanics \cite{EPR35}, nowadays has many
applications, but no explanation.

We are not in position to comment this
phenomenon, and confine ourself instead to less involved {\it
Bell's approach} \cite{Bell65}. Henceforth we completely disregard
the nonlocality, and turn  to quantum correlations per se.

\subhead2.2.~Bell's paradox\endsubhead
Let  $X_i$, $i\in I$ be observables of quantum system $S$, that is
Hermitian operators $X_i\in \Cal G$ from Lie algebra of the
dynamic symmetry group $G$. According to quantum paradigm actual
measurement of $X_i$  in state $\psi$ produces random quantity
$x_i$, determined by expectations of all functions $f(x_i)$
$$\langle f(x_i)\rangle=\langle\psi|f(X_i)|\psi\rangle$$
(the moments $\langle x_i^n\rangle$ are usually enough). If for
some set of indices $J\subset I$ observables $X_j$, $j\in J$
commute, then the random quantities $x_J=\{x_i\mid i\in J\}$ have
{\it joint distribution} given by
$$\langle f(x_J) \rangle=\langle \psi| f(X_J)|\psi\rangle,$$
where $f(x_J)$ is a function of $x_j, j\in J$. The so called
``classical realism" postulates existence of a hidden joint
distribution of {\it all\,} variables, commuting or not. To test
it we have to solve the following problem.
\proclaim{2.2.1.~Marginal problem}
Under which conditions a system of marginal distributions of
$x_J$, $J\subset I$ can be extended to a joint distribution of all
$x_I$?
\endproclaim
This is a question about  existence of a ``body" (= probability
density) in $\Bbb R^I$ with given projections onto some coordinate
subspaces $\Bbb R^J$, $J\subset I$.

Note that univariant margins $x_i$ are always compatible (one can
take joint distribution of $x_i$ as {\it independent\,}
quantities). The following inequality is necessary for consistency
of bivariant margins $x_{ij}=(x_i,x_j)$
$$\Bbb D(x_i)+\Bbb D(x_j)+\Bbb
D(x_k)+2\text{Cov}(x_i,x_j)+2\text{Cov}(x_j,x_k)+2\text{Cov}(x_k,x_i)\ge
0,$$ since  LHS is equal to $\Bbb D(x_i+x_j+x_k)$, {\it provided}
a joint distribution of $x_i,x_j,x_k$ exists. This is a simplest
prototype of Bell's inequalities.

\subsubhead 2.2.1.1.~Remark\endsubsubhead
The marginal problem has a long history, starting from works by
W.~Hoefling in Germany (1940), and a bit later by Frech\'e  in
France. Springer Verlag published collected papers of Hoefling in
1994. Three conferences on the subject held in the last decade
\cite{Marg91, Marg96, Marg97}. None of the participants ever
mentioned Bell's problem, and apparently none of physicists was
aware about these activities. This is a disturbing example of a
split between mathematics and physics.

\subhead{2.3.~Ansatz for testing ``classical realism"}\endsubhead
The random quantity $x_i$, $i\in I$ assumes values in
$\Lambda_i=\operatorname{Spec}X_i$. For $J\subset I$ put
$\Lambda_J=\prod_{j\in J}
\Lambda_j$ and consider functions on $\Lambda=\Lambda_I$ of the
form
$$F(\lambda)=\sum_{X_J\text{ commute}}f_J(\lambda_J),
\quad\lambda\in\Lambda,\tag2.1$$
where \roster
\item"" $J\subset I$ corresponds to {\it commuting\,} sets
of operators $X_j, j\in J$,
\item"" $f_J$ is a real function on $\Lambda_J$,
\item"" $\lambda_J$ is projection of $\lambda\in\Lambda_I$ onto
$\Lambda_J$.
\endroster
Such function $F$, by commutativity of $X_J$, unambiguously
determines Hermitian operator
$$F(X)=\sum_{X_J\text{ commute}}f_J(X_J).\tag2.2$$
\proclaim{2.3.1.~Theorem} State $\psi$ is consistent with classical realism
iff
$$F(\lambda)\ge 0\Rightarrow \langle\psi|F(X)|\psi\rangle\ge 0\tag2.3$$
for all functions $F$ of form (2.1).\endproclaim
\proclaim{2.3.2.~Corollary} Every state is compatible
with classical realism iff
$$F(\lambda)\ge 0\Rightarrow F(X)\ge 0
\tag 2.4$$
for all functions $F$ of form (2.1).
\endproclaim
The proof of the theorem is based on the above considerations and
Kellerer's criterion \cite{Kell64} for solvability of the marginal
problem.
\subsubhead2.3.3.~Remark\endsubsubhead
The set of nonnegative functions $F$ of type (2.1) forms a convex
cone $\Cal K$, which will be called {\it Kellerer's cone}. It is
enough to check (2.3) only for {\it extremal functions} $F\ge 0$
from $\Cal K$, i.e. for those which aren't positive combinations
of others. The corresponding {\it Bell's inequality\,}
$\langle\psi|F(X)|\psi\rangle\ge0$ is also said to be {\it
extremal\,} (it can't be deduced from the others).  The extremal
functions generate {\it edges\,} of the Kellerer's cone $\Cal K$.

\subsubhead2.3.4.~Example\endsubsubhead
Let's consider a system of  two particles $a$ and $b$. The dynamic
symmetry group in this case is $\text{SU}(2)\times\text{SU}(2)$,
and the state space is tensor product $\Bbb H_a\otimes\Bbb H_b$ of
spin spaces of the particles. Let $A_i$ and $B_j$ be spin
projection operators for particles $a$ and $b$ onto directions $i$
and $j$. Operators $A_i$, $B_j$ commute, and for spin 1/2 with two
measurement per site  Kellerer's cone $\Cal K$ is given by
$$F(a_1,a_2,b_1,b_2)=f_{11}(a_1,b_1)+f_{12}(a_1,b_2)+f_{21}(a_2,b_1)+f_{22}(a_2,b_2)
\ge0,$$
where $a_i,b_j=\pm1$ are eigenvalues of $A_i$ and $B_j$.
All the edges of this cone can be obtained from
Clauser-Horn-Shimony-Holt function \cite{CHSH69}
$$a_1b_1+a_2b_1+a_2b_2-a_1b_2+2\ge0$$
by permutation of particles $a\leftrightarrow b$ and switching
eigenvalues $a_i\mapsto \pm a_i$, and $b_j\mapsto \pm b_j$. So we
have essentially one Bell's type inequality for testing
``classical realism"
$$\langle\psi|A_1B_1|\psi\rangle+
\langle\psi|A_2B_1|\psi\rangle+
\langle\psi|A_2B_2|\psi\rangle-
\langle\psi|A_1B_2|\psi\rangle+2\ge0.\tag CHSH$$

\subsubhead 2.3.5.~Remark\endsubsubhead
Finding of vertices or edges is a typical {\it linear
programming\,} problem. Each time I decide to run my computer
overnight, it finds a couple of new extremal Bell's inequalities
for three particles system of spin 1/2. This amounts altogether to
27 nonequivalent classes, including five given in \cite{WerWol01}.
\footnote{The title of this interesting paper may be misleading,
since the authors confine themselves on functions of special form,
which do not exhaust the whole Kellerer's cone. } The list is
probably still incomplete.

Notice that Scarani and Gisin \cite{ScaGis01} relate violation of
Bell's inequalities to security of quantum communication.
In the core of a conventional security system lies  a ``hard
problem", like prime decomposition of an integer $N\gg 1$, that
complexity presumably grows faster then any power of $N$, while
checking of a given solution takes only polynomial time.
Currently, however, there is not a single problem, for which such
widely expected behavior has been rigorously proven.  This is a
one million dollars Millennium Problem
\footnote{Among 6 others, including {\it Riemann Conjecture} and {\it Quantum
Yang-Mills Theory}.} of Clay Mathematical Institute \cite{Cook}.
Quantum computers may drastically change the very notion of
complexity \cite{Shor97}. See also \cite{Pitow89} on complexity of
Bell's type problems.

\proclaim{2.3.6.~Theorem} An irreducible quantum system with dynamic
group $G$ of rank at least two is incompatible with classical
realism.\endproclaim
The {\it rank} of group $G$ is a maximal number of linear
independent commuting operators (=observables) in its Lie algebra
$\Cal G$. For example,
$\operatorname{rk}\operatorname{SU}(n)=n-1$. A system of rank one
can't violates ``classical realism", since one dimensional margins
are always consistent.

A group of a greater rank, contains either $\text{SU(2)}\times
\text{SU}(2)$ or $\text{SU}(3)$. The first case amounts to widely
known violation of Bell's inequalities in two particles systems.
Below is a typical example of a nonclassical behavior in
$\text{SU}(3)$. Recall, that this is {\it chromodynamic group} of
internal states of hadrons. It has another physical incarnation as
{\it polarization group\,} of a massive quantum vector field.
\subsubhead 2.3.7.~Pentagonal inequality\endsubsubhead
Let's consider a cyclic quintuplet of orthogonal states
$$e_i\in\Bbb H,\quad e_i\perp e_{i+1}, \quad i \mod 5\tag2.5$$
in standard three dimensional  representation of $\text{SU}(3)$,
and denote by $S_i=1-2|e_i\rangle\langle e_i|$ {\it reflection
operator} in a mirror  orthogonal to $e_i$. Operators $S_i$,
$S_{i+1}$ commute, and the corresponding Kellerer's cone consists
of functions of the form
$$f_{12}(s_1,s_2)+f_{23}(s_2,s_3)+f_{34}(s_3,s_4)+
f_{45}(s_4,s_5)+f_{51}(s_5,s_1)\ge 0$$ where $s_i$ assumes values
$\pm 1=\text{Spec}\,S_i$. Here is an example of such function
$$s_1s_2+s_2s_3+s_3s_4+s_4s_5+s_5s_1+3\ge 0.\tag2.6$$
Indeed, each summand  $s_is_{i+1}$ is equal to $\pm 1$, while
their product is $+1$. Hence there exists at least one positive
summand $s_is_{i+1}=+1$, and (2.6) follows.

One can show that (2.6) is an extremal function from the
Kellerer's cone, and all such functions  can be obtained from this
one by switching the eigenvalues $s_i\mapsto \pm s_{i}$. Applying
Theorem 3.3.1 we get extremal {\it pentagonal inequality}
$$\langle\psi|S_1S_2|\psi\rangle+\langle\psi|S_2S_3|\psi\rangle+
\langle\psi|S_3S_4|\psi\rangle+
\langle\psi|S_4S_5|\psi\rangle+\langle\psi|S_5S_1|\psi\rangle+
3\ge 0,\tag2.7$$ for testing ``classical realism". One can put it
in geometric form as follows
$$\sum_i \cos^2\alpha_i\le 2,\quad
\alpha_i=\widehat{\psi e}_i.$$
This inequality fails, for example, for a {\it regular}
configuration of vectors $e_i$, and $\psi$ directed along its axis
of symmetry (of order 5). In this case
$$\sum_i \cos^2\alpha_i=\frac{5\cos\pi/5}{1+\cos{\pi/5}}=2.236067.$$
In a smaller extent violation of  the pentagonal  inequality is
almost inevitable in all settings: for every configuration (2.5)
with no collinear vectors, operator $\sum_iS_iS_{i+1}$ has an
eigenvalue $\lambda<-3$, and the corresponding eigenstate $\psi$
breaks classical law (2.7).

Notice that in this example all states are {\it coherent}, and
{\it none} of them is compatible with ``classical realism".

\subsubhead2.3.8.~ Summary\endsubsubhead\roster
\item  ``Classical realism"
fails whenever it is virtually possible (Theorem 2.3.6).
\item It may fail for {\it all\,} states, including coherent ones (see
$\text{n}^\circ$ 2.3.7).
\item A state may be manifestly nonclassical, even if Bell's
approach fails to detect this (see below $\text{n}^\circ$ 2.5.3).
\item Failure of ``classical realism" is a basic fact of quantum mechanics, in
noway specific for entanglement.
\item
Bell's inequalities is a marginal problem indeed. I would say they
lead into a dead end. See however $\text{nn}^\circ$ 2.8, 2.5.4,
2.3.5.
\endroster

\subhead 2.4.~Extremal property of completely entangled states\endsubhead
In the previous section  we have seen  how illusive may be
connection between ``classical realism" and entanglement. Instead
of this ambiguous relation we put forward an {\it extremal
property\,} of a completely entangled state, which can be checked
in all known instances, namely the maximality of its  total
variance:
$${\Bbb D}(\psi):=\sum_i
\langle\psi|X_i^2|\psi\rangle-\langle\psi|X_i|\psi\rangle^2=\max.\tag2.8$$
One can see from equation (1.9)
$$\Bbb D(\psi)=\langle\lambda,\lambda+\rho\rangle-
\sum_i\langle\psi|X_i|\psi\rangle^2$$
that the maximum is attained for state $\psi$ with zero
expectation of all observables
$$\langle\psi|X|\psi\rangle=0,\quad \forall X\in \Cal G,\tag2.9$$
and the maximum itself is equal to Casimir
$$\max_\psi\,\Bbb D(\psi)=\langle\lambda,\lambda+\rho\rangle,\tag2.10$$
Notice  that (2.8) is opposite to the property of {\it coherent
states}, for which the total variance is minimal and equal to
$\langle\lambda,\rho\rangle$ (Theorem 1.3.1). Therefore
generically we have inequality
$$ \langle\lambda,\rho\rangle\le\Bbb D(\psi)\le\langle\lambda,
\lambda+\rho\rangle. \tag2.11$$
\subsubhead{2.4.1.~Remark}\endsubsubhead There is a minor
discrepancy between conditions (2.8) and (2.9). They are
equivalent, {\it provided\,} there exists at least one state with
zero average of all observables. We'll call a system {\it
degenerate} if it has no such states. There are very few
degenerate systems consisting of one component, i.e. with {\it
simple} dynamic group \cite{VinPop92}
\roster\item $n$-dimensional representations of
$\text{SU}(n)$ and $\text{Sp}(n)$.
\item For odd $n$ representation of $\text{SU}(n)$ in
space of skew-symmetric bilinear forms.
\item A halfspinor representation of dimension 16 of
$\text{Spin}(10)$.
\endroster
There are many more such composite systems, and their
classification is also known due to M.~Sato and T.~Kimura
\cite{SatKim77}.
It tells which simple quantum systems can {\it not} be completely
entangled into a composite one. See $\text{n}^\circ$ 2.5.2 for
examples.

\subhead 2.5.~Formal definition and examples\endsubhead
In what follows we assume (2.9), rather then (2.8), as a  formal
definition of a completely entangled state.
\proclaim{2.5.1.~Definition} State $\psi\in \Bbb H$ is said  to be
{ completely entangled} if all observables  $X\in
\Cal G$ have zero expectation in state $\psi$
$$\langle\psi|X|\psi\rangle=0,\quad \forall X\in \Cal G.\tag2.9$$
\endproclaim
Notice, that property (2.9) is $G$-invariant, i.e. the dynamic
group transforms completely entangled state $\psi$ into completely
entangled one $g\psi$, $g\in G$.

Recall also, that the total variance of a completely entangled
state is {\it maximal}, as opposed  to a coherent state, for which
the variance is {\it minimal}.
The total variance is a natural measures of quantum fluctuations
in a system. Therefore one can informally think about coherent
states
as {\it closest to classical}, and completely entangled ones as
{\it manifestly nonclassical}, see $\text{nn}^\circ$ 2.5.3, 2.5.4
for examples.
By this reason purely quantum effects, such as {\it nonlocality},
or violation of {\it classical realism} are most likely to happen
for a completely entangled state.

All the states unanimously recognized as completely entangled
conform with this definition, see examples 2.5.2 and conjecture
2.5.6  below.
But the main argument in its favor comes from equation (2.9),
which is mathematically meaningful, and connects entanglement to
{\it Geometric Invariant Theory} to be discussed in the next
section.

\subsubhead 2.5.2.~Completely entangled states in composite systems\endsubsubhead
 Let's consider composite  system
$$\Bbb H=\bigotimes_{i=1}^N \Bbb H_i\tag2.10$$ with components of dimension
$n_i$ and dynamic group $G_i=\text{SU}(n_i)$. This scheme includes
$N$ {\it qubits system\,} of particles of spin 1/2. Choose
orthonormal basis $e_i$ in $\Bbb H_i$ and  arrange components of
tensor
$$\psi=\sum
\psi_{\alpha_1\alpha_2\ldots\alpha_N}e_1^{\alpha_1}\otimes e_2^{\alpha_2}\otimes \cdots
\otimes e_N^{\alpha_N}$$ into $N$ dimensional matrix $[\psi]$.
Applying to $\psi$ criterion (2.9) one can deduce
\proclaim{2.5.2.1.~Prorposition} State $\psi\in\Bbb H$ is
completely entangled iff parallel slices of its matrix $[\psi]$
are orthogonal and have the same norm.
\endproclaim
\proclaim{2.5.2.2.~Corollary} Composite system (2.10) admits a
completely entangled state iff information {capacities\,}
$\delta_i=\log n_i$, $n_i=\dim\Bbb H_i$ of the components satisfy
polygonal inequalities
$$\delta_i\le\sum_{j(\ne i)}\delta_j.$$
\endproclaim
The inequality follows from linear independence of the orthogonal
parallel slices, which implies $n_j\le n_1n_2\cdots
\widehat{n_j}\cdots n_N$.

\subsubhead 2.5.2.3.~Examples\endsubsubhead
i) For completely entangled state $\psi\in\Bbb H_1\otimes \Bbb
H_2$ in a two components system, the matrix $[\psi]$ has
 format  $m\times n$, $m=\dim\Bbb H_1$, $n=\dim\Bbb
H_2$ with orthogonal rows and columns of the same norm
$1/\sqrt{m}$ and $1/\sqrt{n}$ respectively. This is possible only
if $n=m$, and in this case $[\psi]$ is proportional to a {\it
unitary\,} matrix. This implies that completely entangled state is
unique
$$\psi=\frac1{\sqrt{n}}\sum_ie^i\otimes f^i,\tag EPR$$
up to action of dynamic group $\text{SU}(n)\times\text{SU}(n)$.
For $n=2$ it is known  as EPR or Bell state.

 \quad ii) Similarly in three qubits system there exists unique completely
entangled state
$$\psi=\frac1{\sqrt2}(e_1\otimes e_1\otimes e_1+e_2\otimes e_2\otimes
e_2),\tag GHZ$$ up to action of dynamic group
$\text{SU}(2)\times\text{SU}(2)\times\text{SU}(2)$. This is well
known Greenberg-Horn-Zeilinger state \cite{GHZ89,CHSZ90}.

\quad iii)
The previous two {\it riggid\,} examples are actually exceptional.
For $N>3$ completely entangled $N$ qubits state, modulo action of
the dynamic group, depends on $2^N-3N-1$ complex parameters. The
structure of this {\it moduli space} is not known neither in $N$
qubits setting, nor for a generic  three components system. See
$\text{n}^\circ$~2.5.3 for description of a similar moduli space
of spin entangled states.

\quad iv)
For composite system (2.10)  coherent state $\psi$ is just
decomposable tensors
$\psi=\psi_1\otimes\psi_2\otimes\cdots\otimes\psi_N$ (see no.
1.2.1). Such states for a long time where treated as {\it
completely disentangled}.

\subsubhead 2.5.3.~Completely entangled spin states\endsubsubhead
Let's consider completely entangled state $\psi\in\Bbb H_j$ of a
system of spin $j$.
According to Definition 2.5.1 this means that
average spin projection onto {\it every\,} direction is zero.
This certainly can't happens for $j=1/2$, since in this case each
state has definite spin projection $1/2$ onto some direction. But
for $j\ge 1$ such states do exist. For example, one can take
$\psi=|0\rangle$ for integer $j$, and in general
$$\psi=\frac1{\sqrt{2}}(|+j\rangle+|-j\rangle).$$
Up to a rotation this is the only possibility for $j=1$ or $3/2$.
All completely entangled states of arbitrary spin $j$ can be
constructed as follows \cite{Kly94}.
\proclaim{2.5.3.1.~Ansatz}
Start with a configuration of $2j$ unit vectors $p_i\in\Bbb S^2$
with zero sum (one can visualize it as a closed $2j$-gon with unit
sides in $\Bbb R^3$). Then take their images $\zeta_i\in\Bbb C$
under stereographic projection $\pi:\Bbb S^2\rightarrow \Bbb C$
and expand  the product
$$\prod_{i}(z-\zeta_i)=\sum_{\mu=-j}^{\mu=j}
a_\mu{{2j}\choose{j+\mu}}z^{j+\mu},$$ to end up with completely
entangled state
$$\psi=\sum_{\mu=-j}^{\mu=j}a_\mu{{2j}\choose{j+\mu}}^{1/2}|\mu\rangle,$$
possibly non normalized.
\endproclaim

To sum up: completely entangled states of spin $j$ are
parameterized by closed {\it polygonal strings} in $\Bbb R^3$ of
length $2j$. Evolution and decay of the states may be viewed as
evolution and decay of the string. This is a typical example of
description of coherent states from a perspective of Geometric
Invariant Theory.

Every such state is {\it manifestly nonclassical\,}, since average
projection of moment onto any direction is zero, while  the
standard deviation $\sqrt{j(j+1)}$ {\it exceeds\,} maximum of the
projection $j$.
This kind of nonclassical behavior can't be detected by Bell's
approach, which needs at least two independent commuting
observables, see $\text{n}^\circ$ 2.3.6. But in no way it  is less
nonclassical then EPR.\footnote{Well, except the nonlocality,
which has nothing to do with Bell's inequalities, and remains
enigmatic anyway.}
\subsubhead{2.5.4.~Entangled states in other systems}
\endsubsubhead The previous arguments can be literally extended
onto arbitrary system, using inequality
$\langle\lambda,\lambda\rangle<\langle\lambda,\lambda+\rho\rangle$
instead of $j^2< j(j+1)$:
\proclaim{2.5.4.1.~Claim} Every completely entangled state is
manifestly nonclassical.
\endproclaim
It is easily seen that zero weight vector $\psi$, that is a vector
annihilated by Cartan subalgebra (see $\text{n}^\circ$ 2.8.1), is
always completely entangled. For spin group $\text{SU}(2)$ this
amounts to state $|0\rangle$ with zero spin projection.  For
chromodynamic group $\text{SU}(3)$ this includes hadrons composed
of equal number of all three quarks $u,d,s$ (antiquark is counted
with coefficient -1). For example $\pi^0$ is an entangled state in
octet (=adjoint representation) of spin $0$ mesons, while
$\pi^\pm$ are coherent states in this octet. Big quantum
fluctuations in entangled state may be responsible for instability
of $\pi^0$, which life time nine orders smaller then $\pi^\pm$.
\subsubhead{2.5.5.~Remark}\endsubsubhead
Product $\psi=\psi_1\otimes\psi_2$ of two completely entangled
states is a completely entangled state, although very untypical
one.
For example completely entangled state of two particles of spin
$\ge 1$ may decay into two components, each being entangled onto
itself. The degenerated representations listed in $n^\circ$ 2.4.1
may be used as building blocks for {\it stable systems}, for which
such decay is forbidden.


We close this section with the following conjecture, motivated by
Theorem 2.3.6 and the previous remark.
\proclaim{2.5.6.~Conjecture}
Indecomposable completely entangled state of a system with dynamic
symmetry group of rank at least two is incompatible with classical
realism.
\endproclaim
This can be checked in many cases, but general proof is still
missing. The conjecture is primary designed to convert the
perplexed.

\subhead 2.6.~Kempf-Ness unitary trick and GIT stability\endsubhead
The extremal property (2.8-9) of a completely entangled state is
closely related to concept of  stability in {\it Geometric
Invariant Theory\,} (GIT). The later emerges from the classical,
mostly algebraic, invariant theory of 19th century enhanced with
innovating geometric insight by D.~Hilbert. Later on it was
transformed by D.~Mumford into a powerful universal formalism,
which infinite dimensional version is familiar  to physicists from
gauge theory. The third edition of his book \cite{MFK94} includes
a bibliography about 1000 titles.

Vector $\psi\in\Bbb H$ is said to be {\it semistable} if it can be
separated from zero by a $G$-invariant function $I$, that is
$I(\psi)\ne I(0)$.
Invariant function $I(g\psi)=I(\psi)$, $g\in G$ is just   a {\it
conservation law} or an {\it integral} of the system. We expect
the invariant $I$ to be {\it holomorphic}, in which case it
retains the invariance with respect to complexified group $G^c$.
Notice that Hermitian metric $\langle\psi|\psi\rangle$ is a
$G$-invariant, but not a holomorphic function.

Nonvanishing invariant $I(\psi)\ne 0$ prevents $\psi$ from falling
to zero under action of the complexified group $G^c$. This implies
existence of a nonzero vector $\psi_0=g\psi$, $g\in G^c$ of
minimal length, {\it provided\,} complex orbit $G^c\psi$ of $\psi$
is closed. In the last case state $\psi$ is said to be {\it
stable}.\footnote{In GIT stable state supposed to have at most
finite symmetry group, while condition (2.9) ensures only that its
dimension is as small as possible. We'll not pay much attention to
the distinction between stability and semi\-stability. }
from $g\psi$, $g\in G^c$ as a limit. The following theorem
\cite{KemNes78} identify the minimal vector with a completely
entangled state.
\proclaim{2.6.1.~Kempf-Ness unitary trick}  Orbit $G^c\psi$ is closed iff
it contains   vector $\psi_0=g\psi$ of minimal length. Then the
minimal vector is unique up to (unitary) action of $G$, and can be
defined by  equation
$$\langle\psi|X|\psi\rangle=0,\quad X\in \Cal G.\tag2.9$$
\endproclaim
\proclaim{2.6.2.~Corollary}
Every stable vector belongs to a complex orbit of a completely
entangled state.
\endproclaim
This is a crucial observation for our
approach, which unveils that an adequate mathematics hidden behind
entanglement is Geometric Invariant Theory.

\subsubhead{2.6.3.~Example}\endsubsubhead
Let's consider completely entangled state $\psi_0\in\Bbb H_j$ of a
particle of spin $j$ (see $\text{n}^\circ$ 2.5.3). In moving frame
it takes form $\psi=g\psi_0$ for some $g\in G^c=\text{SL}(2,\Bbb
C)$ (see Example 1.2.3). Notice that matrix
$g=g(\Psi_0)\in\text{SL}(2,\Bbb C)$ depends on the whole wave
function $\Psi_0$ of the particle \cite{PST02}, i.e. a state
vector in an irreducible representation of Lorentz group
$\text{SL}(2,\Bbb C)$, see for details \cite{Wig39}.
 Nobody believes that Lorentz transformation can
completely destroy an entangled state. Therefore, the set of
(partially) entangled states must be closed under action of the
complexified group $G^c$, hence by Corollary 2.6.2  it includes
all stable states. By logical and technical reasons semistable
states also must be included.

This example suggests  equivalence
between two apparently very different concepts.
\proclaim{2.6.4.~Definition} Entangled state $\psi$
is just a semistable vector, that is it can be separated from zero
$I(\psi)\ne I(0)$ by some holomorphic $G$-invariant function $I$.
\endproclaim
Below we consider a number of other examples in support of
conformity and significance of this formal definition.

\subsubhead{2.6.5.~Invariants of a composite system}\endsubsubhead
Let us return to the settings of $\text{n}^\circ$ 2.5.2 and
consider composite system
$$\Bbb H=\bigotimes_{i=1}^N \Bbb H_i\tag2.10$$ with dynamic group of
$i$-th component $\text{SU}(\Bbb H_i)$.
\subsubhead2.6.5.1.~Two component system\endsubsubhead
As we know from Corollary 2.5.2.2 system $\Bbb H_1\otimes\Bbb H_2$
with components of dimensions $m,n$ can't be entangled, except
$n=m$. Hence for $m\ne n$ there are no nontrivial invariants. For
$m=n$ there exists unique basic invariant $I=\det[\psi]$, where
$[\psi]$ is a matrix of tensor $\psi\in
\Bbb H_1\otimes\Bbb H_2$. In two qubits setting $n=2$ this
invariant separates entangled states $\det[\psi]\ne 0$ from
coherent  ones $\det[\psi]=0\, \Leftrightarrow\,\psi=\psi_1\otimes
\psi_2$ (see Example 2.5.2.3.iv).
\subsubhead 2.6.5.2.~Hyperdeterminant\endsubsubhead
For  general system $N$ component system (2.10) under certain
conditions there exists a similar invariant,
called {\it hypereterminant} $\text{Det}[\psi]$ of $N$ dimensional
matrix $[\psi]$, introduced by Gelfand, Kapranov, and Zelevinsky
\cite{GKZ94}. It is nontrivial only if projective dimensions
$\overline{n}_i=\dim\Bbb H_i-1$ satisfy {\it polygonal
inequalities}
$$\overline{n}_i\le\sum_{j(\ne i)}\overline{n}_j.\tag 2.11$$
For $N=2$ this condition confines us to square matrices, where
$\text{Det\,}[\psi]=\det[\psi]$. The hyperdeterminant is invariant
under elementary transformations of parallel slices, and shares
many other properties of convensional determinant. For matrix  of
format $2\times2\times2$  it looks as follows
$$\aligned\text{Det\,}A&=(a_{000}^2a_{111}^2+a_{001}^2a_{110}^2+a_{010}^2a_{101}^2+a_{011}^2a_{100}^2)\\
&-2(a_{000}a_{001}a_{110}a_{111}+a_{000}a_{010}a_{101}a_{111}+a_{000}a_{011}a_{100}a_{111}\\
&\quad +a_{001}a_{010}a_{101}a_{110}+a_{001}a_{011}a_{110}a_{100}+a_{010}a_{011}a_{101}a_{100})\\
&+4(a_{000}a_{011}a_{101}a_{110}+a_{001}a_{010}a_{100}a_{111}).\endaligned\tag2.12$$
Every such matrix can be diagonalized by elementary slice
transformations. Therefore three qubits state is either coherent
$\psi=\psi_1\otimes\psi_2\otimes\psi_3$, if
$\text{Det\,}[\psi]=0$, or entangled, if $\text{Det\,}[\psi]\ne0$.
In the later case it can be transformed by complex dynamic group
into diagonal GHZ state $\text{n}^\circ$ 2.5.2.3.ii).
\proclaim{2.6.5.3.~Conjecture} Invariants of  $N$ qubits system
$$\Bbb H_I=\bigotimes_{i\in I}\Bbb H_i,\quad \dim\Bbb H_i=2$$ are generated by
 hyperdeterminants of $\Bbb H_I=\Bbb
H_{I_1}\otimes\Bbb H_{I_2}\otimes\cdots\otimes\Bbb H_{I_k}$  of
format $2^{|I_1|}\times2^{|I_2|}\times\cdots\times 2^{|I_k|}$ for
all decompositions of $I$ into disjoint components $I_\alpha$.
\endproclaim
For  binary tensor of valency four $\psi_{ijkl}$ the
hyperdeterminants are $\text{Det\,}[\psi_{ijkl}]$ of format
$2\times2\times2\times2$, and three conventional $4\times4$
determinants like $\det[\psi_{ij|kl}]$. If one of these
hyperdeterminants is nonzero, then $\psi$ is entangled. The
conjecture claims the inverse.
\subsubhead{2.6.6.~Invariants of spin states}\endsubsubhead
Invariants of spin $j$ representation $\Bbb H_j$ of $\text{SU}(2)$
is a classical subject, known as {\it Binary Quantics}. Recall,
that a standard model for spin $j$ representation $\Bbb H_j$ is
the space of {\it binary forms}
$$f(x,y)=\sum_{\mu=-j}^ja_\mu{2j\choose j+\mu}x^{j+\mu}y^{j-\mu}$$
of degree $2j$ in which $\text{SU}(2)$ acts by unitary
transformations of $(x,y)$. In this model state $|\mu\rangle$ with
spin projection $\mu$ corresponds to monomial
$$|\mu\rangle={2j\choose j+\mu}^{1/2}x^{j+\mu}y^{j-\mu}.$$
One of commonly known invariants of binary form $f$ is {\it
discriminant} $\Delta(f)$ which vanishes iff the form has multiple
factors in its decomposition into linear factors
$$f(x,y)=\prod_i(\alpha_i x-\beta_i y).$$ By Definition 2.6.4 every state $\psi$
for which $\Delta(\psi)\ne 0$ is entangled. Coherent states from
this point of view  are most degenerate ones. They correspond to
binomials $(\alpha x-\beta y)^{2j}$.

For spin $j=1$ and $3/2$ there are no other independent
invariants, so in these cases $\Delta\ne 0$ is a criterion of
entanglement. For $j=2$ there is an extra invariant, called {\it
catalectican}, which may be defined for all integer $j$
$$C(f)=\det\left(\matrix
a_j&a_{j-1}&\cdots&a_0\\
a_{j-1}&a_{j-2}&\cdots&a_{-1}\\
\cdots&\cdots&\cdots&\cdots\\
a_0&a_{-1}&\cdots&a_{-j}
\endmatrix\right).$$
It has a transparent physical meaning: $C(\psi)=0$ iff state
$\psi$ is a linear combination  of $j$ coherent states ($j+1$ is
always enough). For $j=2$ discriminant $\Delta$ and catalectican
$C$ are all the basic invariants. Hence in this case state $\psi$
is entangled  iff one of them is nonzero.

The complexity of the problem increases drastically with $j$. This
is an amazingly difficult job,  done by classics for $j\le 3$, and
by modern authors for $j=4$ \cite{Shi67}, and partially for
$j=7/2$ \cite{Dix83}.

For all $j$ entangled states can be easily described
geometrically, see $\text{n}^\circ$ 2.8.2.2. The difficulties come
from a perverse desire to put geometry into Procrustean bed of
algebra.

\subhead{2.7.~Density matrix and measure of entanglement}\endsubhead One can
associate with entangled state $\psi\in \Bbb H$ a  density matrix,
or operator, as follows. Let for simplicity  $\psi$  be a {\it
stable} state with no symmetries. Then $\psi$ can be transformed
into a completely entangled state $\psi_0=g\psi$ by element $g\in
G^c$ of complex dynamic group. By Kempf-Ness theorem 2.6.1 such
$g$ is unique up to left multiplication by an element of dynamic
group $G$ acting in $\Bbb H$ by {\it unitary} transformations.
Therefore product $g^\dag g$ is a well defined positive unimodular
operator independent of the above ambiguity in $g$, and we define
{\it density matrix} just by rescaling  it to trace one
$$\rho(\psi)=\frac1{\text{Tr}g^\dag g}g^\dag g.\tag2.13$$
We define also {\it entropy} of entangled state in usual way
$$S(\psi)=-\text{\rm Tr\,}(\rho(\psi)\log\rho(\psi)).\tag2.14$$
Below are some straightforward implications of these definitions.

\subsubhead{2.7.1.~Properties of the density
matrix}\endsubsubhead
\roster
\item  $G$-invariance: $\rho(g\psi)=\rho(\psi)$, $g\in G$.
\item  $\psi$ is completely entangled  $\Leftrightarrow$ $\rho(\psi)$ is a scalar matrix.
\item $\psi$ is completely entangled $\Leftrightarrow$ its entropy
$S(\psi)$ is maximal.
\item The density matrix of entangled state
 $\psi\in\Bbb H_1\otimes\Bbb
H_2\otimes\cdots\otimes\Bbb H_N$ in composite  system splits into
product $\rho(\psi)=\rho_1\otimes\rho_2\otimes\cdots\otimes\rho_N$
of some density matrices of the components.
Hence in this case $S(\psi)=S(\rho_1)+S(\rho_2)+\cdots+S(\rho_N)$.
\endroster

\subsubhead{2.7.2.~Comments} \endsubsubhead
If Hamiltonian $H$ is included in algebra of observables $\Cal G$,
then time evolution $\psi(t)=e^{itH}\psi(0)$ of isolated system is
governed by a one parametric subgroup of $G$. In this case the
first property tells that the density matrix is an {\it integral
of motion}. This is one of the reasons why an external device is
required for cooking an entangled state \cite{CKS02,...}.

The next two properties essentially tell that von Neumann entropy
$S(\psi)$, and density matrix $\rho(\psi)$ itself, are natural
{\it measures of entanglement}. However, precise  definition of
the density matrix for entangled states with symmetries is
expectedly more involved, since the symmetries produce
singularities in the orbit space. Because of the symmetries
quantum entropy of a generic $N$ qubits state is well defined only
for $N\ge 4$. This may looks not so bad if compared with classical
entropy which makes sense only for $N\rightarrow\infty$.

Simple systems in which every state has a nontrivial symmetry
group are all known \cite{KPV76, Sch78}. These are exactly the
systems with functionally independent basic invariants. For spin
systems this happens for $j\le 2$.

Kempf--Ness theorem provides another measure of entanglement, not
so sensitive to the symmetries, namely length of minimal vector
$\psi_0=g\psi$ in complex orbit of entangled state $\psi$. For two
components system it looks a bit strange
$$|\psi_0|^2=n\mid\det[\psi]\mid^{2/n},$$
while the density matrix in this case is something like
$[\psi]^\dag[\psi]$ modulo an ambiguity caused by symmetries of
$\psi$.

A precise physical meaning of all these invariants still needs to
be clarified. Notice also that none of the measures of
entanglement is relativistic invariant \cite{PST02}. For a spin
system every entangled state looks as completely entangled in an
appropriate moving frame, see Example 2.6.3.

\subhead{2.8.~Hilbert-Mumford criterion}\endsubhead
Until now we have two means to distinguish entangled states from
others:
\roster
\item
 Produce $\psi=g\psi_0$ from a completely
entangled state $\psi_0$ by complex dynamic transformation $g\in
G^c $.
\item
Find a holomorphic invariant $I$ which separates $\psi$ from zero.
\endroster
Both approaches have some troubles. The first needs a description
of all completely entangled states, and the second one assumes
knowledge of all basic invariants. Getting either of this
prerequisites is a challenge problem, see $\text{nn}^\circ$
2.5.2.1, 2.5.3.1, 2.6.5, 2.6.6.

Hilbert-Mumford criterion \cite{MFK94}, provides a more practical
way for such characterization, using so called {\it stability
inequalities}.  This approach bear a similarity to that of Bell,
especially in the role played by Cartan subalgebras. Although the
nature of stability inequalities is quiet different from that of
Bell, they retain the main idea that entangled states may be
characterized by {\it some} inequalities. One may expect a close
connection between these two subjects.

\subsubhead2.8.1.~Cartan subalgebras\endsubsubhead
By definition {\it Cartatan subslgebra\,} $\Cal C\subset \Cal G$
is a maximal commutative subalgebra. Its dimension $\dim
\Cal C=r$ is equal to the {\it rank\,} of group $G$, see
$\text{n}^\circ$ 2.3.6. A typical example is algebra of diagonal
matrices in Lie algebra of all (skew) Hermitian matrices. Action
of $\Cal C$ splits state space $\Bbb H$ into orthogonal sum of
eigenspaces spanned by eigenvectors $|e_i\rangle$
$$\Bbb H=\bigotimes_i \Bbb C |e_i\rangle, \quad
X|e_i\rangle=\langle\omega_i,X\rangle|e_i\rangle,\quad X\in \Cal
C.\tag2.15$$ Elements $\omega_i\in\Cal C$ are said to be {\it
weights\,} of $\Bbb H$. Let now decompose state $\psi$ over the
eigenbasis
$$\psi=\sum a_i|e_i\rangle\tag2.16$$
and define its $\Cal C$-{\it support\,\,} $\text{Supp}_{\Cal
C\,}\psi\subset\Cal C $ as convex hull of those weights $\omega_i$
for which
$a_i\ne 0$.
\proclaim{2.8.2.~Hibert-Mumford criterion}
State $\psi$ is stable iff zero is an interior  point of $\Cal
C$-support\,\, $\text{\rm Supp}_{\Cal C\,}\psi$ for every Cartan
subalgebra $\Cal C\subset \Cal G$, and semistable iff it is never
outside of the support.\endproclaim

Returning back to entanglement we may spell out this as follows
$$\text{State $\psi$ is entangled}\quad\Longleftrightarrow\quad
0\in \text{Supp}_{\Cal C}\psi,\quad\forall\,\,\Cal C.\tag2.17$$
Moreover if zero is always an {\it internal} point of the support
then state $\psi$ is stable with finite symmetry group. In the
last case the density matrix (2.13) is well defined.
\subsubhead2.8.2.1.~Entangled states in $N$ qubits
system\endsubsubhead Applying Hilbert-Mumford criterion (2.17) to
$N$-qubit state
$$\psi=\sum
\psi_{s_1s_2\ldots s_N}|s_1\rangle\otimes |s_2\rangle\otimes \cdots
\otimes |s_N\rangle,\quad s_i=\pm$$
we find out that $\psi$ is entangled iff zero  is contained in the
convex hull of points $\{(s_1,s_2,\ldots,s_N)\in \Bbb R^N\mid
\psi_{s_1s_2\ldots s_N}\ne 0\}$ whichever directions are used for
spin projections $s_i$.
\subsubhead 2.8.2.2.~Entangled states in spin system\endsubsubhead
In this case Hilbert-Mumford criterion tells that state $\psi\in
\Bbb H_j$ of spin $j$ is {\it not\,} entangled iff $\psi$ is a
linear combination of states with {\it positive} spin projections
onto some direction.
$$\psi=\sum_{0\le j-\mu<j} a_\mu|\mu\rangle$$


\head 3.~Conclusion\endhead
Group theoretical approach is inevitably more kinematic then
dynamic. But this shortage may turn into an advantage in searching
for the very basic concepts.


\refstyle A

\enddocument
\newpage

\subhead
Add to paper
\endsubhead
\subsubhead 1.  Dynamic symmetry group\endsubsubhead
Lie groups and Lie algebras. Complexification. Cartan subalgebras,
rank. Adjoint representations and roots. Representations and
dominant weights weights.

\subsubhead 2. Entangled states as stable vectors\endsubsubhead
Stability as a nondegeneracy condition. Characterization stable
states by moment map (Hilbert-Mumford criterion). Characterization
of entangled states by a nonvanishing invariant. Examples: Two,
three, and four particle entangled states. Completely entangled
states as multidimensional ``orthogonal" matrices (p. 13). Special
case of 3-dimensional matrices (p.13b-14). Necessary condition for
existence of entangled state in this case (logarithmic polynomial
inequality, see end of p13b). Multidimensional determinants, and
other invariants (p. 17+). Conjectural structure of invariants in
$N$-particle state (see p. 15). Binary forms and entangled spin
states, Catalictican and decompositions over coherent states
(p.15,17,18). Vanishing of three dimensional determinant. Finding
invariants in general is a challenging problem. Completely
entangled states of a system of particles of arbitrary spin as
multidimensional matrices with orthogonal slices (see p. 13, 15).
 ``Bad" representations with no stable vectors (see p.11, 18). Gauge theory as
example of dynamic group formalism, and as a model for
entanglement.
\subsubhead 3. Kempf--Ness unitary trick\endsubsubhead
Relation with moment map. Completely entangled states. Measure of
entanglement (p. 27). Entropy of entangled state, see also
\cite{PST02} on {\it dependence} of entanglement from Lorentz
transformation, cf. p.1,no.X, p.2 no.3.

\subsubhead 4.  Local criterion of entanglement\endsubsubhead
Vanishing of  average of all infinitesimal operators of the
dynamic group. Examples.

\subsubhead 5. Maximal variance property\endsubsubhead
$$\sum\Bbb D_\psi(X_\alpha)=<\lambda,\lambda+\rho>,$$ see p. 9.

\subsubhead 6. Examples\endsubsubhead
\roster
\item"i)" EPR case.
\item"ii)" One particle state of arbitrary spin. Pentagon
experiment.
\item"iii)"
Entangled states of a particle and antiparticle.
\item"iv)" Three particles states. Cubic determinant, and other invariants.
\item"v)" Entangled $\operatorname{SU}(3)$ states. Muon  as
entangled state of quark and antiquark. Proton as entangled state
of three quarks.
\item"vi)" Quantum crystal as an example of entangled state.
\endroster

Because of the above connection between entangled and coherent
states, we begin  with an exposition the basic facts from the
theory of coherent states \cite{Perel86}.

Entanglement, its two striking  features:
\subsubhead 2.1.1.  Nonlocality
\endsubsubhead
\subsubhead 2.1.2. Violation of ``classical realism"\endsubsubhead
Bell's inequalities as a marginal problem for a system of
partially commuting operators (p.7+1/2, p. 28). Overview of the
marginal problem (p. 19+). Pitowsky's theorem: The Bell's problem
is equivalent to \$1000.000 Clay Institute Millennium Prize
problem "P versus NP" \cite{Pitow89, Cook}, but it is not so
surprising since even $\text{Windows}^\copyright$ Minesweeper game
do the same (there is a survey of Pitowsky's results in my
notebook 15, p.158). All non-quantum cryptography is based on {\it
unproved} inequality $P\ne NP$.

Examples:
\roster
\item"i)" Two spin 1/2 particles.
Unique completely entangled state (EPR, or Bell), one
 inequality CHSH (Clauser, Horn, Shimony, Holt). See p. 11,12,
\item"ii)" Three spin 1/2 particles. Unique completely entangled
state GHZ (Greenberger, Horn, Zeilinger) five {\it homogeneous}
inequalities and {\it dozens}
 inhomogeneous ones. see p.13 for arbitrary number of particles.
\item"iii)" Violation of
classical realism in {\it coherent} $SU(n)$ states, $n\ge 3$.
Pentagonal experiment.
\item"iv)" Entangled states of a particle of spin $j$. (see p. 15)
\item"v)" Entangled states of an $n$ particle system of arbitrary
spin and polygon spaces. see p. 13.
\endroster

\subsubhead 2.1.3.  Why this properties fail to produce
a meaningful definition\endsubsubhead
\roster
\item"i)"  Nonlocality refers to a composite spatially separated systems.
Therefor it has only asymptotic meaning, and can be applied only
to a limited class of systems, not, for example, to a localized
system. Accepting nonlocality as the essence of EPR paradox, and
the most striking manifestation of entanglement, we reject it as a
decisive property because of this lack of universality.
\item"ii)"  Violation of ``classical realism", on the
contrary, is one of the most common effects in quantum mechanics,
which is just a manifestation of interference of wave functions.
Because of this lack of specificity it can't serve as a basis for
a substantial theory. Furthermore, not all entangled states
violate Bell's type inequalities \cite{Gisin96, Peres96}.
\endroster


\head
1. Coherent states
\endhead

\subhead
1.1. Short history
\endsubhead
Coherent states where introduced by Schr\"odinger
\cite{Schr\"o26}, but was almost completely forgotten until
Glauber \cite{Glaub63} rediscovered them in connection with laser
emission. Later on Perelomov \cite{Perel86} put them into an
adequate context of action the dynamic symmetry group on the state
space of a system. We'll use the same approach for entanglement,
and therefore recall some basic facts about coherent states.

\subhead
1.2. Weil-Heisenberg algebra
\endsubhead

\subhead
1.3. Coherent states as  complex orbit of vacuum
\endsubhead
 General definition. Main  properties:
\subsubhead
1.3.1. Minimal uncertainty
\endsubsubhead
$$\Delta p\Delta
q=\hbar/2$$ (see p. 8).
\subsubhead
1.3.2.  Minimal variance
\endsubsubhead
$$\sum_i\Bbb D_\psi(X_i)=<\lambda,\rho>$$ (
see p.9+) where $X_i$ is an orthonormal basis of Lie algebra
$\frak G$ of the dynamic symmetry group $G$. We suppose the state
space $\Bbb H$ of the system to be   irreducible with highest
weight $\gamma$, and $\rho$ is the halfsum of positive roots.

\subsubhead 1.3.3. Coherent states are most close to classical
ones\endsubsubhead This statement, suggested by the previous two,
being not free of an ambiguity, expresses an important heuristic
principle.

\subsubhead 1.3.4. Overcompleteness\endsubsubhead
This is  an example of mistreating of coherent states as
``overcomplete coordinate system", rather then an important
physical notion.

\subsubhead 1.3.5.  Summary
\endsubsubhead
There is no coherent state without dynamic symmetry group, which
should be chosen and fixed beforehand. Coherent state theory is a
physical equivalent of Kirillov--Kostant--Sourio  {\it orbit
method} in representation theory \cite{Kiril}.

\head
2. Entanglement: Search for definition
\endhead
\subhead 2.0. Short history\endsubhead
The very term ``entanglement" originates from the famous
Schr\"odinger's ``cat paradox"
\footnote{As BBC puts it: {\it In quantum mechanic it is not so
easy to be or not to be.}}
 paper \cite{Schr\"o35}, which in
turn was inspired by the no less celebrated
Einstein--Podolsky--Rosen gedanken experiment \cite{EPR35}. While
the authors stress a nonlocal instant connection between the
particles involved in the experiment, J.~Bell  was the first to
note that it is inconsistent with ``classical realism"
\cite{Bell65}. Since then Bell's inequalities are produced in
industrial scale \cite{GHSZ90, GHZ89, WerWol02,...}

\subhead 2.1. EPR paradox\endsubhead
Entanglement, its two striking  features:
\subsubhead 2.1.1.  Nonlocality
\endsubsubhead
\subsubhead 2.1.2. Violation of ``classical realism"\endsubsubhead
Bell's inequalities as a marginal problem for a system of
partially commuting operators (p. 7+1/2, p. 28). Overview of the
marginal problem (p. 19+). Pitowsky's theorem: The Bell's problem
is equivalent to \$1000.000 Clay Institute Millennium Prize
problem "P versus NP" \cite{Pitow89}, \cite{Cook}.

Examples:
\roster
\item"i)" Two spin 1/2 particles.
Unique completely entangled state (EPR, or Bell), one
 inequality CHSH (Clauser, Horn, Shimony, Holt).
\item"ii)" Three spin 1/2 particles. Unique completely entangled
state GHZ (Greenberger, Horn, Zeilinger) five {\it homogeneous}
inequalities and {\it dozens}
 inhomogeneous ones.
\item"iii)" Violation of
classical realism in {\it coherent} $SU(n)$ states, $n\ge 3$.
Pentagon experiment.
\item"iv)" Entangled states of a particle of spin $j$. (see p. 15)
\endroster

\subsubhead 2.1.3.  Why this properties fail to produce
a meaningful definition\endsubsubhead
\roster
\item"i)"  Nonlocality refers to a composite spatially separated systems.
Therefor it has only asymptotic meaning, and can be applied only
to a limited class of systems, not, for example, to a localized
system. Accepting nonlocality as the essence of EPR paradox, and
the most striking manifestation of entanglement, we reject it as a
decisive property because of this lack of universality.
\item"ii)"  Violation of ``classical realism", on the
contrary, is one of the most common effects in quantum mechanics,
which is just a manifestation of interference of wave functions.
Because of this lack of specificity it can't serve as a basis for
a substantial theory. Besides, not all entangled states violate
Bell's type inequalities \cite{Gisin96, Peres96}.
\endroster
\subhead
2.2. Definition and properties of entangled states
\endsubhead
\subsubhead 2.2.1.  Dynamic symmetry group\endsubsubhead
Lie groups and Lie algebras. Complexification. Cartan subalgebras,
rank. Adjoint representations and roots. Representations and
dominant weights weights.

\subsubhead 2.1.2. Entangled states as stable vectors\endsubsubhead
Stability as a nondegeneracy condition. Characterization stable
states by moment map (Hilbert-Mumford criterion). Characterization
of entangled states by a nonvanishing invariant. Examples: Two,
three, and four particle entangled states. Completely entangled
states as multidimensional ``orthogonal" matrices (p. 13). Special
case of 3-dimensional matrices (p.13b-14). Necessary condition for
existence of entangled state in this case (logarithmic polynomial
inequality, see end of p13b). Multidimensional determinants, and
other invariants (p. 17+). Conjectural structure of invariants in
$N$-particle state (see p. 15). Finding invariants in general is a
challenging problem. Completely entangled states of a system of
particles of arbitrary spin as multidimensional matrices with
orthogonal slices (see p. 13, 15).
 ``Bad" representations with no stable vectors (see p.11, 18). Gauge theory as
example of dynamic group formalism, and as a model for
entanglement.
\subsubhead 2.1.3. Kempf--Ness unitary trick\endsubsubhead
Relation with moment map. Completely entangled states. Measure of
entanglement (p. 27). Entropy of entangled state, see also
\cite{PST02} on {\it dependence} of entanglement from Lorentz
transformation, cf. p.1,no.X, p.2 no.3.

\subsubhead 2.1.4.  Local criterion of entanglement\endsubsubhead
Vanishing of  average of all infinitesimal operators of the
dynamic group. Examples.

\subsubhead 2.1.5. Maximal variance property\endsubsubhead
$$\sum\Bbb D_\psi(X_\alpha)=<\lambda,\lambda+\rho>,$$ see p. 9.

\subsubhead 2.1.6. Examples\endsubsubhead
\roster
\item"i)" EPR case.
\item"ii)" One particle state of arbitrary spin. Pentagon
experiment.
\item"iii)"
Entangled states of a particle and antiparticle.
\item"iv)" Three particles states. Cubic determinant, and other invariants.
\item"v)" Entangled $\operatorname{SU}(3)$ states. Muon  as
entangled state of quark and antiquark. Proton as entangled state
of three quarks.
\item"vi)" Quantum crystal as an example of entangled state.
\endroster


\enddocument